\documentclass[12pt, reqno]{amsart}
\usepackage{amsmath, amsthm, amscd, amsfonts, amssymb, graphicx, color}
\usepackage[bookmarksnumbered, colorlinks, plainpages]{hyperref}

\newtheorem{theorem}{Theorem}[section]
\newtheorem{lemma}[theorem]{Lemma}

\theoremstyle{definition}
\newtheorem{definition}[theorem]{Definition}
\newtheorem{example}[theorem]{Example}

\theoremstyle{remark}
\newtheorem{remark}[theorem]{Remark}
\numberwithin{equation}{section}

\newcommand{\tgh}[1]{\tanh(x_{#1})}
\newcommand{\ath}{\mathrm{atanh}}
\newcommand{\athc}[1]{\ath \frac{#1}{c}}
\newcommand{\norm}[1]{\|#1\|}
\newcommand{\normv}[1]{\norm{\mathbf{#1}}}

\begin{document}
\setcounter{page}{1}

\title[Two-dimensional Einstein numbers]{Two-dimensional Einstein numbers and associativity}

\author[T. Gregor, J. Halu\v{s}ka]{Tom\'{a}\v{s} Gregor$^1$$^{*}$ and J\'an Halu\v{s}ka$^2$}

\address{$^{1}$ Mathematical  Institute, Slovak Academy of  Sciences,  branch Ko\v{s}ice.}
\email{\textcolor[rgb]{0.00,0.00,0.84}{gregor@saske.sk}}

\address{$^{2}$ Mathematical  Institute, Slovak Academy of  Sciences,  branch Ko\v{s}ice.}
\email{\textcolor[rgb]{0.00,0.00,0.84}{jhaluska@saske.sk}}




\date{Received: August 14, 2013; Revised: yyyyyy; Accepted: zzzzzz.
\newline \indent $^{*}$ Corresponding author.}

\begin{abstract}
In this paper, we deal with generalizations of real Einstein numbers to various spaces and dimensions. We search operations and their properties in generalized settings. Especially, we are interested in the generalized operation of hyperbolic addition to more-dimensional spaces, which is associative and commutative. We extend the theory to some abstract spaces, especially to Hilbert-like ones. Further, we bring two different two-dimensional generalizations of Einstein numbers and study properties of new-defined operations -- mainly associativity, commutativity, and distributive laws.
\end{abstract} 

\maketitle

\section{Introduction}
 
The attraction to the operation $\oplus$, the so called hyperbolic velocity addition, was took by Einstein in his famous 1905 paper \cite{Einstein1905} about Special theory of relativity. Therefore it is  also known as the \textit{Einstein operation}.  

The  mathematical motivation to study of Einstein numbers can be explained as follows. Let $u, v \in \mathbb{R}$, be two individual velocities of two moving bodies along a line in one direction and  $0 \leq u < c$, $0 \leq  u < c$, where $c$ denotes the speed of light, the maximal speed in vacuum. If the moving bodies are  related, then the "relativistic" addition $ u \oplus v$ of velocities $u$ and $v$, expressed with the \textit{Einstein number}, has to be less than $c$ as well. The set of all Einstein numbers has to be closed with respect the operation $\oplus$. Here are two expressions of this operation: one in the absolute units [m/s],
\begin{equation}
u \oplus v = \frac{u+v}{1+\frac{uv}{c^2}}, \quad 0 \leq u, v, u \oplus v < c,   \label{EN}\end{equation}
or, relatively to the speed of light, normalized with $c$, no physical units, i.e.,
$$ \frac{u \oplus v}{c} = \frac{\frac{u}{c} + \frac{v}{c}}{1+ \frac{u}{c}\frac{v}{c}}, \quad 0 \leq \frac{u}{c}, \frac{v}{c}, \frac{u \oplus v}{c} <1.$$

Although we are  physically not able to produce or reach the speed of light $c$, \textit{this speed physically really  exists} and, therefore, it is  natural to include the speed of light into the Einstein numbers theory as its own and proper element, $c$ is an Einstein number per definition. We have:

\begin{definition}\label{EREN} (Definition of real Einstein numbers)  Let $c > 0$. Let  $\mathbb{R}_\infty = \mathbb{R} \cup \{ \infty \}$  be the  projective real line with one additional compactification element, denoted as $\infty$. Under the set of all \textit{real Einstein numbers} we understand the system  $\mathbb{E}_{(-c,c]}^\infty = ( (-c, c] = \varphi(\mathbb{R}_\infty), \oplus)$, where $\varphi: \mathbb{R}_\infty \to (-c, c]$ is a bijective function  given as follows: $\varphi(\infty)= c$ and  $\varphi(v) = c \tanh(v)$ for every $v \in \mathbb{R}$. Denote the restriction of $\mathbb{E}_{(-c,c]}^\infty$ to $(-c,c)$ by $\mathbb{E}_{(-c,c)}$.
\end{definition}

\begin{remark}
Davis in his 1940 book, cf. \cite{Davis1940}, and Baker in the 1954 paper, cf. \cite{Baker1954}, coined the term Einstein numbers. Davis was interested in the properties of the element $c$ in $\mathbb{E}_{(-c,c]}^\infty$, while Baker in the creating of the operation multiplication to obtain a field ($c$ does not belong to this field).
Another theory directly bounded with Einstein numbers is a theory of Ungar. This theory follows  aims of hyperbolic geometry and physics. The  generalization of Einstein operation in his work is non-associative and non-commutative, cf. \cite{Ungar2007}.
\end{remark}

\section{Monoid structures}
\label{monoid}
Recall that a system $\mathbb{E} = (E,  \oplus)$ consisting of a set $E$ and a binary operation $\oplus$ on $E \times E$,  is called to be a \textit{monoid} if
\begin{enumerate} 
\item[(i)]    $u, v \in E \implies u \oplus v \in E$;
\item[(ii)] the operation $\oplus: E \times E \to E$ is associative;
\item[(iii)]   there exists $e \in E$ (the \textit{identity element}) such that $e \oplus u = u \oplus e =u$ for all $u \in E$. 
\end{enumerate}

Note that, in general,  a monoid need not to be commutative.

The following lemma is an useful tool that makes possible to extend the given binary operations and to prove its properties.

\begin{lemma}\label{lm_izomorf}
Let $A,B$ be sets and the binary operation $\ast$ on $A$ be associative (commutative, have identity element, etc.). If $F: A \to B$ is a bijective function, then the binary operation $\circledast$ on $B$ defined by $x\circledast y=F\!\left(F^{-1}(x) \ast F^{-1}(y)\right)$, $x,y\in B$, is associative (commutative, have identity element, etc.)
\end{lemma}
\proof
Let $\ast$ be the associative binary operation on $A$. Then for $x,y,z \in B$
$$\begin{array}{rcl}
(x\circledast y)\circledast z &=& F\!\left(F^{-1}(x) \ast F^{-1}(y)\right) \circledast z
  \medskip \\ &=&
  F\!\left[F^{-1}\!\left[F\!\left(F^{-1}(x) \ast F^{-1}(y)\right)\right] \ast F^{-1}(z)\right]
  \medskip \\ &=&
  F\!\left[\left(F^{-1}(x) \ast F^{-1}(y)\right) \ast F^{-1}(z)\right]
  \medskip \\ &=&
  F\!\left[F^{-1}(x) \ast \left( F^{-1}(y) \ast F^{-1}(z)\right)\right]
  \medskip \\ &=&
  F\!\left[F^{-1}(x) \ast F^{-1}\!\left[F\!\left( F^{-1}(y) \ast F^{-1}(z)\right)\right]\right]
  \medskip \\ &=&
  x \circledast F\!\left( F^{-1}(y) \ast F^{-1}(z)\right)
  \medskip \\ &=&
  x \circledast (y \circledast z),
\end{array}$$
thus $\circledast$ is associative. Similarly it can be proved that if $\ast$ is commutative, so $\circledast$ is commutative too, if $\ast$ have identity element $e$, so $\circledast$ have identity element $F(e)$, and that many other properties are transferred from the operation $\ast$ to the operation $\circledast$.
\qed

\begin{lemma}\label{lm_monoid}
Let $c >0$. Let the operation $\oplus: [0,c] \times [0,c]$ is described with the formula \eqref{EN}. Then the system $\mathbb{E}_{[0,c]} = \left([0, c], \oplus \right)$ is a monoid.
\end{lemma}
\proof 
It is followed from the properties of the hyperbolic tangent function, that if $u, v \in [0, c]$, then 
$$u \oplus v=\frac{u+v}{1+\frac{uv}{c^2}}=c\tanh\!\left(\!\ath\frac{u}{c}+\ath\frac{v}{c}\right)$$
The function $\varphi: [0,\infty] \to [0,c]$, $\varphi(u)=c \tanh u$, $u\in[0,\infty)$, $\varphi(\infty)=c$ is bijective and its inverse is $\varphi^{-1}: [0,c] \to [0,\infty]$, $\varphi^{-1}(u)=\ath \frac{u}{c}$, $u\in[0,c)$, $\varphi^{-1}(c)=\infty$. Thus from the previous lemma the operation $\oplus$ is closed and associative (additionally, is also commutative) on $[0, c]$, since the operation of classical $+$ is associative (and commutative) on $[0,\infty]$, if we define $x+\infty=\infty+x=\infty$ for all $x\in[0,\infty]$.

\begin{lemma}
Let $c>0$. The monoid  $\mathbb{E}_{[0,c]}$ can be extended to the system of real Einstein numbers $\mathbb{E}_{(-c,c]}^\infty$ via the formula (\ref{EN}).
\end{lemma}
\proof
A proof formally copies the previous one, we extend the isomorphism $\varphi$ on the set $(-c,c]$. 
\qed

We collect the distinguished elementary properties of Einstein numbers into the following lemma.
\begin{lemma}
(a) The restricted Einstein numbers $\mathbb{E}_{(-c,c)}$ are a group, but Einstein numbers $\mathbb{E}_{(-c,c]}^\infty$ are not. 

(b) Einstein numbers $\mathbb{E}_{(-c,c]}^\infty$ (and thus also $\mathbb{E}_{(-c,c)}$) are linearly ordered. 

(c) Einstein numbers $\mathbb{E}_{(-c,c]}^\infty$ are a compact space, but $\mathbb{E}_{(-c,c)}$ are not (they are only locally compact space). 
\end{lemma}

\proof
(a) The set $\mathbb{E}_{(-c,c)}$ is isomorphic to the real line under the isomorphism $\varphi(u)=c \tanh u$, thus they are an additive group. However, Einstein numbers are not an additive group; the element $c$ has no inverse, since $u \oplus c = c$ for all $u \in (-c,c]$. 

(b) The linear ordering is a specific property of Einstein numbers, $-c < u \leq c$, $u \in \mathbb{E}_c$.  In general, a one-point compactification damages order structures.
	
(c) The basis of the standard topology which corresponds to one-point compactification consists of the sets of the form $(a,b)$ and $(-c,a) \cup (b,c]$, where $-c<a<b<c$.
\qed	

Here is a review of monoid structures of Einstein numbers:

\begin{lemma}  Let $c >0$. Einstein numbers $((-c, c], \oplus)$ contains the following substructures which have restricted domains but are closed with respect to (restricted) operation $\oplus$:

(i) $(\{0,c\}, \oplus)$ is the 0-1 addition table;

(ii) $([0, c], \oplus)$ is a monoid;

(iii) $((-c,0], \oplus)$ is a monoid;

(iv) $(-c, c)$ is isomorphic to $\mathbb{R}$;

(v) Einstein numbers $((-c, c], \oplus)$ is a glue (in the sense of operation $\oplus$) of two monoids.
  \end{lemma}
\proof

(i) $$ \begin{array}{c|cc}
\oplus & 0 & c \\ \hline 0 & 0 & c \\ c & c & c
\end{array}$$

(ii) see Lemma~\ref{lm_monoid};

(iii) analogously (ii);

(iv) cf. \cite{Baker1954}, the real field is isomorphic to field $((-c,c),\oplus,\odot)$ under the transformation $c \tanh(a)$, where 
\begin{equation}\label{Ein_plus_krat}
\begin{array}{c}
 u\oplus v=c \tanh\!\left(\!\ath\dfrac{u}{c}+\ath\dfrac{v}{c}\right)=\dfrac{u+v}{1+\frac{u v}{c^2}}\medskip \\
u\odot v=c\tanh\!\left[\left(\!\ath\dfrac{u}{c}\right)\left( \ath\dfrac{v}{c}\right)\right].
\end{array}
\end{equation}

(v) we have to prove that if $u \in (-c, 0]$ and $v \in [0, c]$, then $u \oplus v$ has a sense. If $v\neq c$, then (iv). If $v = c$, then $u\oplus v=u\oplus c=c$.
\qed

\begin{remark} 
If $c \to \infty$, then the $c$-ball expands to infinity and Einstein addition reduces to the "normal" addition, i.e. Einstein addition is reduced to the Newtonian addition.
\end{remark}

\section{A theory in the three dimensional Euclidean space}
This section consists of some comments to generalizations of Einstein numbers given in works of Ungar, cf. e.g. \cite{Ungar2007}.

Ungar studied a generalized Einstein operation $\oplus$ in its three dimensional generalization defined on 
$$ \mathbb{R}_c^3 = \{\mathbf{v} \in \mathbb{R} \mid \|\mathbf{v}\| < c\}$$
 of the all relativistically admissible velocities.
Although this operation is widely accepted by physicists, from the mathematical viewpoint has no good algebraic properties - it is non-associative, non-commutative the so called 
gyrogroup (nothing saying about the operation of multiplication).

Recall briefly basics about this
\textit{extension of hyperbolic velocity addition} to the three dimensional Euclidean space.
The extended hyperbolic velocity addition to the three dimensional Euclidean space, $\bigoplus: \mathbb{R}^3_c \times \mathbb{R}^3_c \to
\mathbb{R}^3_c$, of velocities is given by the equation
\begin{equation}\label{E3D}
\mathbf{u} \bigoplus \mathbf{v} = \frac{1}{1 + \frac{\langle
\mathbf{u} \mid \mathbf{v}\rangle }{c^2}} \left\{ \mathbf{u} +
\mathbf{v} + \frac{1}{c^2} \frac{\gamma_{\mathbf{u}}} {1 +
\gamma_{\mathbf{u}}}(\mathbf{u} \times (\mathbf{u} \times
\mathbf{v})) \right\}
\end{equation}
for all $\mathbf{u},\mathbf{v} \in \mathbb{R}^3_c$, where $\langle\mathbf{u} \mid
\mathbf{v}\rangle$ and $\mathbf{u} \times \mathbf{v}$ are the
inner product and the vector product that the ball
$\mathbb{R}^3_c$ inherits from space $\mathbb{R}^3$, and where
$\gamma_{\mathbf{u}}$ is the \textit{gamma factor}
$$\gamma_{\mathbf{u}} = \frac{1}{\sqrt{1-
\frac{\| \mathbf{u} \|^2}{c^2}}}$$
in the $c$-ball, $\| \mathbf{u}\|^2=\langle\mathbf{u} \mid \mathbf{u}\rangle$.

The gamma factor is related to Einstein addition by the identity
$$\gamma_{\mathbf{u} \bigoplus \mathbf{v}}
= \gamma_\mathbf{u} \gamma_\mathbf{v}\! \left(\! 1 + \frac{\langle
\mathbf{u} \mid \mathbf{v} \rangle}{c^2}\right)$$
and provides the link between Einstein's special
theory of relativity and hyperbolic Lobachevsky geometry, cf.
\cite{Ungar2007}.

This way described generalization of Einstein numbers plays an important role in the theory of
Bergman space $\mathcal{A}^2$ (the space of all analytical
functions on the unit ball $\mathbb{D}$), where it has a form
\begin{equation}\label{00}  w \bigoplus z
 =\frac{1}{1+\langle w \mid z \rangle} \left\{\left[ 1 +
\frac{\langle w \mid z \rangle}{\|w\|^2} \left( 1 -  \sqrt{1-
\|w\|^2}\right)\right] w +  \left[\sqrt{1- \|w\|^2}\right] z
\right\}. \end{equation}  For more details, cf. Rudin,
\cite{Rudin}, Sec. 2.29,
and Zhu, \cite{Zhu}, Eq. (1.2), where the operation $\bigoplus$ is
an involutive automorphism, the self map $\varphi_w: \mathbb{D}
\to \mathbb{D}$  and
$$\begin{array}{l}\varphi_w(z)
\\ \quad\quad
 =\frac{1}{1 - \langle w \mid z \rangle} \left\{\left[ 1 -
\frac{\langle w \mid z \rangle}{\|w\|^2} \left( 1 -  \sqrt{1-
\|w\|^2}\right)\right] y -  \left[\sqrt{1- \|w\|^2}\right] z
\right\}
\medskip\\ \quad \quad
 =\frac{1}{1+\langle w \mid (-z) \rangle} \left\{\left[ 1 +
\frac{\langle w \mid (-z) \rangle}{\|w\|^2} \left( 1 -  \sqrt{1-
\|w\|^2}\right)\right] y +  \left[\sqrt{1- \|w\|^2}\right](-z)
\right\}
\medskip \\ \quad \quad 
= w \bigoplus (-z),
\end{array}$$
for every $w \in \mathbb{D}$; $\langle\cdot \mid \cdot \rangle$ denotes the scalar product in $\mathbb{C}$. This (two dimensional) variant of
the operation $\bigoplus$ can be generalized to $\mathbb{C}^n$, were
$n$ is an arbitrary natural number.

Expansion of the operation $\bigoplus$ to arbitrary inner vector spaces is done with the following way.
 Owing to the well-known Lagrange vector identity, Davis--Snider, \cite{Davis},
 \begin{equation}\label{Lagr}\mathbf{A} \times (\mathbf{B} \times
 \mathbf{C})
 = \mathbf{B} \langle\mathbf{A} \mid \mathbf{C}\rangle
 - \mathbf{C} \langle\mathbf{A} \mid \mathbf{B}\rangle,
\end{equation}
where $\mathbf{A}, \mathbf{B}, \mathbf{C} \in \mathbb{R}^3$,
Einstein addition \eqref{E3D} can also be rewritten as following,
cf. \cite{Fock64, Sexl},
\begin{equation}\label{EIP}
\mathbf{u} \bigoplus \mathbf{v} = \frac{1}{1 +
\frac{\langle\mathbf{u} \mid \mathbf{v}\rangle}{c^2}} \left[
\mathbf{u} + \frac{1}{\gamma_{\mathbf{u}}}\mathbf{v} +
\frac{1}{c^2} \frac{\gamma_{\mathbf{u}}} {1 +
\gamma_{\mathbf{u}}}\langle\mathbf{u} \mid \mathbf{v}\rangle
\mathbf{u} \right].
\end{equation}

The crucial trick of the generalization of the hyperbolic velocity addition is the representation \eqref{EIP} of the representation \eqref{E3D}. Underline that the formula \eqref{E3D} holds only in the three dimensional Euclidean space. But via the formula \eqref{Lagr} we may extend addition on every real inner product space $\mathbb{V}$ (of arbitrary dimension), cf. \cite{Ungar2007}, Definition~1 and~2. This means for finite dimensions 4,5, \dots, and also for spaces non equipped with the finite base. Rather the special case is the dimension 2 (and 4, 8, 16, too) which case can be obtain via the way from dimension 3 but also directly from the original definition of Einstein numbers.

\section{Case of the "deformed" linear normed spaces}

Let us come back to the Section \ref{monoid}, where we expressed real Einstein numbers as an isomorphic image of real numbers under the isomorphism $\varphi(u)=c \tanh u$. We can expand this idea to more general spaces. What about "deformed" linear normed spaces without loss of associativity and commutativity? The first idea is to extend the support of the isomorphism $\varphi$ from the interval $(-c,c)$ to the $c$-ball in some linear normed space.

Let $\mathbb{V}$ be a vector space over the field of real numbers equipped with a norm $\norm{\cdot}$ and an operation of addition $+$. Let us denote the $c$-ball in $\mathbb{V}$ 
$$ \mathbb{V}_c = \{\mathbf{v} \in \mathbb{V} \mid \normv{v} < c\}.$$ Let us define the function $\phi: \mathbb{V} \to \mathbb{V}_c$ as follows
\begin{equation}\label{izo_LNS}
\phi(\mathbf{u})=\left\{
\begin{array}{ll}
\dfrac{c \tanh\normv{u}}{\normv{u}}\mathbf{u} &  \mathrm{if \ } \mathbf{u}\neq \mathbf{0}, \\
0 & \mathrm{if \ } \mathbf{u}= \mathbf{0}.
\end{array}\right.
\end{equation}

\begin{lemma}
The function $\phi$ defined above is a bijection between $\mathbb{V}$ and $\mathbb{V}_c$ and its inverse is
\begin{equation}\label{inv_LNS}\phi^{-1}(\mathbf{u})=\left\{
\begin{array}{ll}
\ath\!\left(\!\dfrac{\normv{u}}{c}\!\right)\dfrac{\mathbf{u}}{\normv{u}} & \mathrm{if \ } \mathbf{u}\neq \mathbf{0}, \\
0 &  \mathrm{if \ }\mathbf{u}= \mathbf{0}.
\end{array}\right.
\end{equation}
\end{lemma}
\proof
First we prove that $\phi$ is an injective function. Let $\mathbf{u},\mathbf{v} \in \mathbb{V}$. If $\phi(\mathbf{u})=\phi(\mathbf{v})=\mathbf{0}$, then $\mathbf{u}=\mathbf{v}=\mathbf{0}$. Let now $\phi(\mathbf{u})=\phi(\mathbf{v})\neq\mathbf{0}$. Then
$$\dfrac{c \tanh\normv{u}}{\normv{u}}\mathbf{u}=\dfrac{c \tanh\normv{v}}{\normv{v}}\mathbf{v},$$
thus vectors $\mathbf{u},\mathbf{v}$ are collinear, i.e. $\mathbf{u}=k \mathbf{e},\mathbf{v}=l \mathbf{e}$ for some scalars $k,l\neq 0$ and a vector $\mathbf{e}\in\mathbb{V}$ with $\normv{e}=1$. We get
$$\dfrac{k \tanh |k|}{|k|}=\dfrac{l \tanh |l|}{|l|}.$$
There holds
$$\dfrac{x \tanh |x|}{|x|}=\tanh x$$
for all $x\in \mathbb{R}\setminus\{0\}$. Function $\tanh x$ is injective, hence $\mathbf{u}=\mathbf{v}$. It implies that $\phi$ is injective.

Now let $\mathbf{v}\in \mathbb{V}_c$. If $\mathbf{v}=0$, then $\phi(\mathbf{0})=\mathbf{v}$. So, let $\mathbf{v}\neq\mathbf{0}$. If we choose
$$\mathbf{u}=\dfrac{\ath\frac{\normv{v}}{c}}{\normv{v}}\mathbf{v},$$
then
$$\phi(\mathbf{u})=\mathbf{v}.$$
So, $\phi$ is a surjective function.

It can be also trivially proved that
$$\phi\!\left(\phi^{-1}\!(\mathbf{u})\right)=\mathbf{u},\quad \phi^{-1}\!\left(\phi(\mathbf{v})\right)=\mathbf{v}$$
for all $\mathbf{u}\in \mathbb{V}_c$ and $\mathbf{v}\in \mathbb{V}$.
\qed

Now we are ready to extend the operation of the hyperbolic tangent addition from the real line to every linear normed vector space over real numbers. To do this we use the isomorphism $\phi$ and Lemma~\ref{lm_izomorf}. Define the binary operation $\oplus_{\mathbb{V}_c}$ in the set $\mathbb{V}_c$ as follows
\begin{equation}\label{plus_LNVS}
\textbf{u}\oplus_{\mathbb{V}_c} \mathbf{v}=c \tanh\!\left( \| A(\mathbf{u},\mathbf{v}) \|\right) \frac{A(\mathbf{u},\mathbf{v})}{\| A(\mathbf{u},\mathbf{v}) \|}
\end{equation}
where
$$A(\mathbf{u},\mathbf{v})=\ath\!\left(\!\frac{\normv{u}}{c}\!\right) \frac{\textbf{u}}{\normv{u}}+\ath\!\left(\!\frac{\normv{v}}{c}\!\right) \frac{\textbf{v}}{\normv{v}}.$$
It is clear that $(\mathbb{V}_c,\oplus_{\mathbb{V}_c})$ is a commutative group.

\begin{theorem}
Let $\mathbb{V}=\mathbb{C}$ be the complex plane equipped with a norm $\norm{z}=\sqrt{a^2+b^2}$, where $z=a+b i\in \mathbb{C}, a \in \mathbb{R}, b \in \mathbb{R}$. Let us denote the $c$-ball in $\mathbb{C}$, $c>0$, as following
$$ \mathbb{C}_c = \{z \in \mathbb{C} \mid \norm{z} < c\}.$$ Then 
\begin{equation}
u\oplus_{\mathbb{C}_c} v=c \tanh\!\left( \| A(u,v) \|\right) \frac{A(u,v)}{\| A(u,v) \|}
\end{equation}
where
$$A(u,v)=\ath\!\left(\!\frac{\norm{u}}{c}\!\right) \frac{u}{\norm{u}}+\ath\!\left(\!\frac{\norm{v}}{c}\!\right) \frac{v}{\norm{v}}$$
and
\begin{equation}
u\odot_{\mathbb{C}_c} v=c \tanh\!\left( \| M(u,v) \|\right) \frac{M(u,v)}{\| M(u,v) \|}
\end{equation}
where
$$M(u,v)=\ath\!\left(\!\frac{\norm{u}}{c}\!\right) \frac{u}{\norm{u}}\cdot\ath\!\left(\!\frac{\norm{v}}{c}\!\right) \frac{v}{\norm{v}}.$$
\end{theorem}

It can be easily checked through isomorphism $\phi$ that $(\mathbb{C}_c\setminus \{ 0\},\odot_{\mathbb{C}_c})$ is a commutative group and also that there hold the distributive laws between the operations of addition and multiplication. Thus $(\mathbb{C}_c,\oplus_{\mathbb{C}_c},\odot_{\mathbb{C}_c})$ is a field isomorphic to the complex numbers $\mathbb{C}$. If we express $u,v\in \mathbb{C}_c\setminus \{0\}$ in polar coordinates, 
$$\mathbf{u}=r e^{i\alpha},\quad \mathbf{v}=s e^{i\beta},\quad 0<r<c, 0<s<c,\; \alpha,\beta \in \mathbb{R},$$
then the operation $\odot_\mathbb{C}$ in the set $\mathbb{C}_c$ is given as follows
$$\mathbf{u} \odot_{\mathbb{C}_c} \mathbf{v} = r e^{i\alpha} \odot_{\mathbb{C}_c} s e^{i\beta} = c \tanh\!\left[\left(\! \ath\frac{r}{c}\right) \left(\ath\frac{s}{c} \right)\right]\cdot e^{i(\alpha+\beta)}.$$

\section{Generalization to more dimensions}
We study another possibilities of extension of isomorphism $\varphi: \mathbb{R} \to \mathbb{R}$ to $\varphi_n: D\subset\mathbb{R}^n \to \mathbb{R}^n$ and define new operations by this isomorphism in this section. To do this we claim the following lemma.

\begin{theorem}\label{izomorf_Rn}
Let $D_1,D_2,\ldots,D_n\subset\mathbb{R}$, $H_0,H_1,\ldots,H_n\subset\mathbb{R}$. The function $f_0:D_1 \to H_0$ and for all indexes $i=1,2,\ldots,n$ $f_i:D_i \to H_i$ are bijective functions, where $f_i(x)\neq 0$ for all $x\in D_i, i=1,2,\ldots,n-1$. Define $F: D=D_1 \times D_2 \times \ldots \times D_n \to F(D)$ by
$$F(x_1,x_2,\ldots,x_n)=\left(f_0(x_1),\frac{f_2(x_2)}{f_1(x_1)},\frac{f_3(x_3)}{f_2(x_2)},\ldots,\frac{f_n(x_n)}{f_{n-1}(x_{n-1})}\right)$$
and put $a=f_1\!\left(f_0^{-1}(x_1)\right)$. If $f_1\!\left(a x_2 x_3\cdot\ldots\cdot x_i\right)\in H_i, i=2,3,\ldots,n$, then
$$F^{-1}\!(x_1,x_2,\ldots,x_n)=
\big(f_0^{-1}(x_1),f_2^{-1}(a x_2),f_3^{-1}(a x_2 x_3),\ldots,f_n^{-1}(a x_2 x_3 \ldots x_n)\big).$$
\end{theorem}
\proof
We have under the above assumptions
$$\begin{array}{l}
F\!\left(F^{-1}\!(x_1,x_2,\ldots,x_n)\right)
 \medskip \\ \qquad =
 F\!\left(f_0^{-1}(x_1),f_2^{-1}(a x_2),\ldots,f_n^{-1}(a x_2 x_3 \ldots x_n)\right)
 \medskip \\ \qquad = \displaystyle{
 \left(f_0(f_0^{-1}(x_1)),\frac{f_2(f_2^{-1}(a x_2))}{f_1(f_0^{-1}(x_1))},\ldots,\frac{f_n(f_n^{-1}(a x_2 x_3 \ldots x_n))}{f_{n-1}(f_{n-1}^{-1}(a x_2 x_3 \ldots x_{n-1}))}\right)}
 \medskip \\ \qquad = \displaystyle{
 \left(x_1,\frac{a x_2}{a},\ldots,\frac{a x_2 x_3 \ldots x_{n-1} x_n}{a x_2 x_3 \ldots x_{n-1}}\right)}=(x_1,x_2,\ldots,x_n)
\end{array}$$
and
$$\begin{array}{l}
F^{-1}\!\!\left(F(x_1,x_2,\ldots,x_n)\right)
  \medskip \\ \qquad = \displaystyle{
  F^{-1}\!\!\left(f_0(x_1),\frac{f_2(x_2)}{f_1(x_1)},\frac{f_3(x_3)}{f_2(x_2)},\ldots,\frac{f_n(x_n)}{f_{n-1}(x_{n-1})}\right)}
  \medskip \\ \qquad = \displaystyle{
  \left[f_0^{-1}\!\left(f_0(x_1)\right), f_2^{-1}\!\left(f_1\!\left(f_0^{-1}\!\left(f_0(x_1)\right)\right) \frac{f_2(x_2)}{f_1(x_1)}\right),\ldots,\right. }
  \medskip \\ \qquad \quad \displaystyle{  
  \left. f_n^{-1}\!\left(f_1\!\left(f_0^{-1}\!\left(f_0(x_1)\right)\right) \frac{f_2(x_2)}{f_1(x_1)} \frac{f_3(x_3)}{f_2(x_2)} \ldots \frac{f_n(x_n)}{f_{n-1}(x_{n-1})}\right)\right]}
  \medskip \\ \qquad = \displaystyle{
  \left[x_1, f_2^{-1}\!\left(f_1(x_1) \frac{f_2(x_2)}{f_1(x_1)}\right),\ldots,
  f_n^{-1}\!\left(f_1(x_1) \frac{f_n(x_n)}{f_1(x_1)}\right)\right]}
  \medskip \\ \qquad =
  \left(x_1, f_2^{-1}\!\left(f_2(x_2)\right),\ldots,
  f_n^{-1}\!\left(f_n(x_n)\right)\right)=(x_1,x_2,\ldots,x_n).
\end{array}$$
\qed

\begin{example}
Let us consider bijective function $F_n: D_n \to H_n$ given by
$$F_n(\mathbf{x})=\left\{
\begin{array}{ll}
\!\!\!\!\Bigg(\!\!\tgh{1},\dfrac{\tgh{2}}{\tgh{1}},\dfrac{\tgh{3}}{\tgh{2}},\ldots,\dfrac{\tgh{n}}{\tgh{n-1}}\Bigg) & \!\!\text{if } x_n>0, 
\medskip \\
\!\!\!\!\Bigg(\!\!\tgh{1},0,\ldots,0\Bigg) & \!\!\text{if } x_n=0,
\end{array}\right.$$
for $\mathbf{x}=(x_1,x_2,\ldots,x_n)\in D_n$, where $$D_n=\left\{(x_1,x_2,\ldots,x_n) \in{\mathbb{R}^n} \mid 0< x_n\leq x_{n-1}\leq \ldots\leq x_1 \right\}\cup \left([0,\infty)\times \{0\}^{n-1}\right)$$ and $$H_n=(0,1)\times(0,1]^{n-1} \cup \left([0,1)\times\{0\}^{n-1}\right).$$ 

By Theorem~\ref{izomorf_Rn},
$$F_n^{-1}\!(\mathbf{x})=\left\{
\begin{array}{ll}
\!\!\!\!\Big(\ath(x_1),\ath(x_1 x_2),\ldots,\ath(x_1 x_2 \ldots x_n)\bigg) & \!\!\text{if } x_n>0, 
\medskip \\
\!\!\!\!\Big(\ath(x_1),0,\ldots,0\bigg) & \!\!\text{if } x_n=0,
\end{array}\right.
$$
for $\mathbf{x}=(x_1,x_2,\ldots,x_n)\in H_n$. Define the operation $\bigoplus$ on $H_n$ by Lemma~\ref{lm_izomorf}, thus
$$\begin{array}{rcl}
\mathbf{x} \bigoplus \mathbf{y} &=& F\!\left(F^{-1}\!(x_1,x_2,\ldots,x_n) + F^{-1}\!(y_1,y_2,\ldots,y_n)\right)
\medskip \\
 &=& 
 \Bigg[\!\tanh\!\left(\ath(x_1)+\ath(y_1)\right)\!,
\medskip \\ & & 
 \dfrac{\tanh\!\left(\ath(x_1 x_2)+\ath(y_1 y_2)\right)}{\tanh\!\left(\ath(x_1)+\ath(y_1)\right)},
\medskip \\ & & 
 \dfrac{\tanh\!\left(\ath(x_1 x_2 x_3)+\ath(y_1 y_2 y_3)\right)}{\tanh\!\left(\ath(x_1 x_2)+\ath(y_1 y_2)\right)},\ldots,
\medskip \\ & & 
 \dfrac{\tanh\!\left(\ath(x_1 \ldots x_n)+\ath(y_1 \ldots y_n)\right)}{\tanh\!\left(\ath(x_1 \ldots x_{n-1})+\ath(y_1 \ldots y_{n-1})\right)}\Bigg]
\medskip \\ 
 &=&
\left(x_1\oplus y_1, \dfrac{x_1 x_2\oplus y_1 y_2}{x_1\oplus y_1}, \dfrac{x_1 x_2 x_3\oplus y_1 y_2 y_3}{x_1 x_2\oplus y_1 y_2},\ldots, \right.
\medskip \\ & &
\left. \dfrac{x_1 x_2 \ldots x_n \oplus y_1 y_2 \ldots y_n}{x_1 x_2 \ldots x_{n-1} \oplus y_1 y_2 \ldots y_{n-1}}\right),
\end{array}$$
if $x_n>0, y_n>0$, otherwise
$$\mathbf{x} \bigoplus \mathbf{y}=\left(x_1\oplus y_1,0, \ldots,0\right),$$
where $\oplus$ is Einstein addition defined by \eqref{EN} with $c=1$. By Lemma~\ref{lm_izomorf}, the operation $\bigoplus$ is associative, commutative, with the identity element $(0, 0, \ldots, 0)$, since the set $D_n$ with usual coordinatewise addition has these properties. Thus $(H_n,\bigoplus)$ is a commutative monoid.
\end{example}

\begin{example}
Let $n=2$ in the previous example. Then we can define the addition in the greater set $D_2^0=D_2\cup\{(0,0)\}$, where $D_2=\left\{(x_1,x_2) \in{\mathbb{R}^2} \mid |x_2|<x_1 \right\}$ is an open cone in the plane, by isomorphism $F_2: D_2^0 \to H_2^0=H_2 \cup \{(0,0)\}$, $H_2=(0,1)\times [0,1)$ given like above
$$F_2(x_1,x_2)=\left\{
\begin{array}{ll}
\!\!\!\!\Bigg(\!\!\tgh{1},\dfrac{\tgh{2}}{\tgh{1}}\Bigg) & \text{if } x_1>0, 
\medskip \\
\!\!\!\left(0,0\right) & \text{if } x_1=0.
\end{array}\right.
$$

Then we obtain the following operation of addition for $\mathbf{x}=(x_1,x_2),\mathbf{y}=(y_1,y_2)\in H_2^0$
$$\mathbf{x} \bigoplus \mathbf{y}=\left\{
\begin{array}{ll}
\!\!\!(0,0) & \text{if } \mathbf{x}=\mathbf{y}=\mathbf{0},
\medskip \\
\!\!\!\!\left(x_1\oplus y_1, \dfrac{x_1 x_2\oplus y_1 y_2}{x_1\oplus y_1}\right) & \text{otherwise}.
\end{array}
\right.$$
Let us notice that the set $D_2^0$ is closed under the multiplication of the so-called \textit{hyperbolic} complex numbers given as follows $(a,b)\cdot(c,d)=(ac+bd,ad+bc)$. We define the next operation by this multiplication and isomorphism $F_2$, namely
$$\begin{array}{rcl}
\mathbf{x} \bigodot \mathbf{y} &=& \Bigg(\!\!\tanh\!\left[\ath(x_1) \ath(y_1)+\ath(x_1 x_2) \ath(y_1 y_2)\right],
\medskip \\ & &
\dfrac{\tanh\!\left[\ath(y_1) \ath(x_1 x_2)+\ath(x_1) \ath(y_1 y_2)\right]}{\tanh\!\left[\ath(x_1) \ath(y_1)+\ath(x_1 x_2) \ath(y_1 y_2)\right]}\Bigg),
\end{array}$$
if $\mathbf{x},\mathbf{y}\in H_2$, and otherwise $\mathbf{x} \bigodot \mathbf{y}=(0,0)$, where $\mathbf{x}=(x_1,x_2),\mathbf{y}=(y_1,y_2)$.
This operation distributes over the operation $\bigoplus$, $(H_2^0,\bigoplus)$ is commutative monoid and $(H_2,\bigodot)$ is a commutative group. Thus the triple $(H_2^0,\bigoplus,\bigodot)$ forms a semi-field with zero, \cite{Vechto}.
\end{example}

\section{Mean-like Einstein numbers}

In this section, we find another and different generalizations of Einstein numbers. This generalization is based on the following assertion, for more details cf. \cite{Hal}.
\begin{lemma}\label{agg}
Let $\zeta: B\to B$ be a bijective function, $(A,\oplus,\odot)$ be a field, and $\boxplus: B \times B \to B$, $\boxdot: A \times B \to B$ be such operations that there holds $p\boxdot\left(x\boxplus y \right)= \left(p\boxdot x\right)\boxplus \left(p\boxdot y\right)$ for all $p\in A, x,y\in B$ and $p\boxdot \left(q \boxdot x\right)=\left(p\odot q\right)\boxdot x$ for all $p,q\in A, x\in B$. Define an operation $\bigoplus: (A\times B)\times (A\times B) \to A\times B$ as follows
\begin{multline*}
  (a_1,b_1) \bigoplus (a_2,b_2)=\Big(a_1 \oplus a_2,   \medskip \\
  \zeta^{-1}\!\!\left[a_1\odot \{a_1\oplus a_2\}^{-1}_A\boxdot \zeta(b_1) \boxplus a_2\odot \{a_1\oplus a_2\}^{-1}_A\boxdot \zeta(b_2) \right]\! \Big),
\end{multline*}
where $c^{-1}_A$ means an inverse element to the element $c$ in the set $A$ with respect to $\odot$. We put $(a_1,b_1) \bigoplus (a_2,b_2)=(0,b)$ for some $b\in B$, if $a_1 \oplus a_2=0$. If $a_1\oplus a_2 \neq 0$, $a_2\oplus a_3 \neq 0$ and $a_1\oplus a_2 \oplus a_3\neq 0$, then
$$\left[(a_1,b_1) \bigoplus (a_2,b_2)\right] \bigoplus (a_3,b_3) = (a_1,b_1) \bigoplus \left[ (a_2,b_2) \bigoplus (a_3,b_3)\right].$$
If moreover $\boxplus$ is a commutative operation, then $(a_1,b_1),(a_2,b_2)$ commute with respect to $\bigoplus$.

\end{lemma}

\begin{remark}
This construction can be easily extended to arbitrary finite number of coordinates.
\end{remark}

\begin{example}
Now let us take operations $\oplus,\boxplus$, resp. $\odot,\boxdot$ as Einstein one-dimensional addition, resp. multiplication defined by hyperbolic tangent isomorphism \eqref{Ein_plus_krat}, with $A=[0,c)$, $B=(-c,c)$. There are satisfied assumptions of Lemma~\ref{agg}. If $\zeta: (-c,c) \to (-c,c)$ is a bijective function, then there exists only function $\eta: \mathbb{R} \to \mathbb{R}$ such that
$$\zeta(x)=c \tanh\!\left[\eta\!\left(\!\athc{x}\right)\right]$$
for all $x\in (-c,c)$. After some manipulations
\begin{multline*}
  (a_1,b_1) \bigoplus (a_2,b_2)=\Bigg(\dfrac{a_1 + a_2}{1+\frac{a_1 a_2}{c^2}} ,  \medskip \\
 c \tanh \eta^{-1}\!\!\left[\dfrac{\eta\!\left(\athc{b_1}\right) \athc{a_1} + \eta\!\left(\athc{b_2}\right) \athc{a_2}}{\athc{a_1} + \athc{a_2}}\right] \Bigg).
\end{multline*}
If $a_1=a_2=0$ we put $(a_1,b_1) \bigoplus (a_2,b_2)=(0,0)$.

\begin{enumerate}
\item[(a)] If $\eta(x)=x$ for all $x\in\mathbb{R}$ we have 
\begin{multline*}
  (a_1,b_1) \bigoplus\nolimits_A (a_2,b_2)=\Bigg(\dfrac{a_1 + a_2}{1+\frac{a_1 a_2}{c^2}},  \medskip \\
  c \tanh \dfrac{\athc{b_1} \athc{a_1} + \athc{b_2} \athc{a_2}}{\athc{a_1} + \athc{a_2}} \Bigg).
\end{multline*}

\item[(b)] If $\eta(x)=1/x$ for all $x\in\mathbb{R}\setminus \{0\}$, $\eta(0)=0$, then 
\begin{equation*} (a_1,b_1) \bigoplus\nolimits_H (a_2,b_2)=\left(\dfrac{a_1 + a_2}{1+\frac{a_1 a_2}{c^2}}, c \tanh \dfrac{\athc{a_1} + \athc{a_2}}{\frac{\athc{a_1}}{\athc{b_1}} + \frac{\athc{a_2}}{\athc{b_2}}} \right).
\end{equation*}
\end{enumerate}

The second coordinate resembles the weighted arithmetic mean in the first case and the weighted harmonic mean in the second case for $b_1,b_2$. Hence the name "mean-like" Einstein numbers. Both operations form a commutative monoid on the set $[0,c) \times (-c,c)$.
\end{example}

\medskip
\noindent \textbf{Acknowledgement.} This paper was supported by Grants VEGA 2/0035/11.

\bibliographystyle{amsplain}

\end{document}